\documentstyle[epsf]{mn}
 \input epsf.sty
\newif\ifAMStwofonts
\ifoldfss

  \ifCUPmtlplainloaded \else
    \NewTextAlphabet{textbfit} {cmbxti10} {}
    \NewTextAlphabet{textbfss} {cmssbx10} {}
    \NewMathAlphabet{mathbfit} {cmbxti10} {} 
    \NewMathAlphabet{mathbfss} {cmssbx10} {} 
  \fi
  \ifAMStwofonts
    \ifCUPmtlplainloaded \else
      \NewSymbolFont{upmath} {eurm10}
      \NewSymbolFont{AMSa} {msam10}
      \NewMathSymbol{\upi}     {0}{upmath}{19}
      \NewMathSymbol{\umu}     {0}{upmath}{16}
      \NewMathSymbol{\upartial}{0}{upmath}{40}
      \NewMathSymbol{\leqslant}{3}{AMSa}{36}
      \NewMathSymbol{\geqslant}{3}{AMSa}{3E}

    \fi
  \fi
\fi 

\ifnfssone
  \newmathalphabet{\mathit}
  \addtoversion{normal}{\mathit}{cmr}{m}{it}
  \addtoversion{bold}{\mathit}{cmr}{bx}{it}
  \newmathalphabet{\mathbfit} 
  \addtoversion{normal}{\mathbfit}{cmr}{bx}{it}
  \addtoversion{bold}{\mathbfit}{cmr}{bx}{it}
  \newmathalphabet{\mathbfss} 
  \addtoversion{normal}{\mathbfss}{cmss}{bx}{n}
  \addtoversion{bold}{\mathbfss}{cmss}{bx}{n}
  \ifAMStwofonts
    \ifCUPmtlplainloaded \else
      \UseAMStwoboldmath
      \makeatletter
      \new@mathgroup\upmath@group
      \define@mathgroup\mv@normal\upmath@group{eur}{m}{n}
      \define@mathgroup\mv@bold\upmath@group{eur}{b}{n}
      \edef\UPM{\hexnumber\upmath@group}
      \new@mathgroup\amsa@group
      \define@mathgroup\mv@normal\amsa@group{msa}{m}{n}
      \define@mathgroup\mv@bold\amsa@group{msa}{m}{n}
      \edef\AMSa{\hexnumber\amsa@group}
      \makeatother
      \mathchardef\upi="0\UPM19
      \mathchardef\umu="0\UPM16
      \mathchardef\upartial="0\UPM40
      \mathchardef\leqslant="3\AMSa36
      \mathchardef\geqslant="3\AMSa3E
    \fi
  \fi
\fi 

\ifnfsstwo
  \DeclareMathAlphabet{\mathbfit}{OT1}{cmr}{bx}{it}
  \SetMathAlphabet\mathbfit{bold}{OT1}{cmr}{bx}{it}
  \DeclareMathAlphabet{\mathbfss}{OT1}{cmss}{bx}{n}
  \SetMathAlphabet\mathbfss{bold}{OT1}{cmss}{bx}{n}
  \ifAMStwofonts
    \ifCUPmtlplainloaded \else
      \DeclareSymbolFont{UPM}{U}{eur}{m}{n}
      \SetSymbolFont{UPM}{bold}{U}{eur}{b}{n}
      \DeclareSymbolFont{AMSa}{U}{msa}{m}{n}
      \DeclareMathSymbol{\upi}{0}{UPM}{"19}
      \DeclareMathSymbol{\umu}{0}{UPM}{"16}
      \DeclareMathSymbol{\upartial}{0}{UPM}{"40}
      \DeclareMathSymbol{\leqslant}{3}{AMSa}{"36}
      \DeclareMathSymbol{\geqslant}{3}{AMSa}{"3E}
    \fi
  \fi
\fi 

\ifCUPmtlplainloaded \else
  \ifAMStwofonts \else 
    \def\upi{\pi}
    \def\umu{\mu}
    \def\upartial{\partial}
  \fi
\fi

\title[Superhumps in Classical Novae]
  {Predicting the Future of Superhumps in Classical Nova Systems}

\author[A. Retter and E.M. Leibowitz]
   {A.~Retter$^*$
   and E.M.~Leibowitz\\
  School of Physics and Astronomy and the Wise Observatory,
Raymond and Beverly Sackler Faculty of Exact Sciences,\\
Tel-Aviv University, Tel Aviv, 69978, Israel\\
   $^*$ email: alon@wise.tau.ac.il\\}

\date{submitted 1997 September 5}
\date{accepted 1998 March 5}
\pagerange{\pageref{firstpage}--\pageref{lastpage}}
\pubyear{1998}

\def\LaTeX{L\kern-.36em\raise.3ex\hbox{a}\kern-.15em
    T\kern-.1667em\lower.7ex\hbox{E}\kern-.125emX}

\begin{document}

\label{firstpage}

\maketitle

\begin{abstract}

Oscillations observed in the light curve of Nova V1974 Cygni 1992 since
summer 1994 have been interpreted as permanent superhumps. From simple
calculations based on the Tidal-Disk Instability model of Osaki, and
assuming that the accretion disc is the dominant optical source in the
binary system, we predict that the nova will evolve to become an SU UMa
system as its brightness declines from its present luminosity by
another 2-3 magnitudes. Linear extrapolation of its current rate of
fading (in magnitude units) puts the time of this phase transition
within the next 2-4 years. Alternatively, the brightness decline will
stop before the nova reaches that level, and the system will continue
to show permanent superhumps in its light curve. It will then be
similar to two other old novae, V603 Aql and CP Pup, that still display
the permanent superhumps phenomenon 79 and 55 years, respectively,
after their eruptions. We suggest that non-magnetic novae with short
orbital periods could be progenitors of permanent superhump systems.

\end{abstract}

\begin{keywords}

novae - stars: individuals: V1974~Cyg - stars: individuals: V603~Aql -
stars: individuals: CP~Pup - stars: evolution - stars: white dwarfs - stars: oscillations - accretion discs

\end{keywords}

\section{Introduction}

\subsection{The (permanent) superhump phenomenon}

Regular superhumps are quasi-periodic oscillations that appear in the
light curve (LC) of the SU UMa subclass of dwarf novae, superimposed on
the superoutburst LC of these systems. The superhump period is a few
percent longer than the orbital period of the system in which it is
observed (laDous 1993). Permanent superhumps (PSH), on the other hand,
appear permanently or most of the time in systems where they prevail,
and do not demand a superoutburst for their emergence. Superhumps are
detected in the LC of cataclysmic variables (CVs) with short orbital
periods ($P_{orbital}$=17-200 min. - Ritter \& Kolb 1998). Stolz \&
Schoembs (1984) found a linear relation between the relative excess of
the superhump period over the orbital one, and the orbital period.

The Tidal-Disc Instability model (for a review see Osaki 1996) is the
commonly accepted explanation of the phenomenon. The superhump
periodicity is the beat of the orbital period of the binary system with
the period of the precession of the accretion disc around the white
dwarf (WD) in the binary co-rotating frame. The difference between
regular superhumps and PSH is understood as a result of the much
larger mass transfer rate in the system showing PSH than in SU UMa
systems. In fact, Osaki's schematic diagram (Fig. 1) states that the
values of only two parameters, the orbital period and the mass transfer
rate, determine the basic differences among the four major classes of
CVs, namely, U Gem, SU UMa, PSH and Nova-like variables.

\begin{figure}

\centerline{\epsfxsize=3.0in\epsfbox{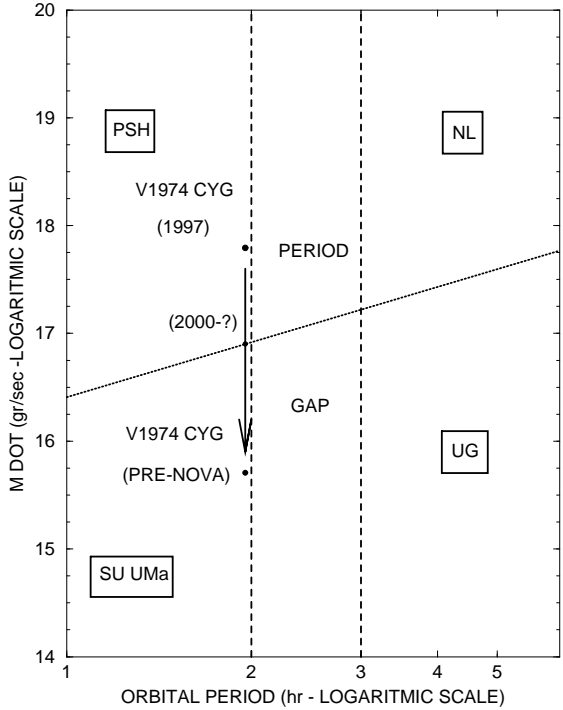}}

\caption{A schematic theoretical diagram based on Osaki (1996), showing
the location on the (Orbital Period, $\dot{M}$) plane of four major
groups of CVs: UG=U Geminorum, SU UMa, PSH=permanent superhumps and
NL=nova like variables. The two dashed vertical lines represent the two
ends of the well known period gap in the period distribution of CVs.
Systems on the right hand side, the UG and NL groups, have accretion
discs that are tidally stable. Systems on the left, PSH and SU UMa, are
tidally unstable. Systems above the dotted tilted line, PSH and NL,
have thermally stable discs. Systems below that line, UG and SU UMa,
have thermally unstable discs. The upper dot represents the present
location of V1974 Cyg in this plane. The arrow indicates the expected
decrease in the $\dot{M}$ value in the future. The lower point is the
location of value of $\dot{M}$, corresponding to the pre-nova magnitude
according to Skiff (1997). If the fading of the nova, in magnitude
units, continues linearly, the arrow will intersect the tilted line
around the year 2000.}

\end{figure}

\subsection{Superhumps in Classical Nova Systems}

Photometric observations of V1974 Cyg (N. Cyg 1992) revealed two
periodic oscillations in the LC of the star (Retter, Ofek \& Leibowitz
1995). While there is an overall agreement that the shorter is the
orbital period of the binary system, two interpretations for the nature
of the second period, which is about 5\% larger than the orbital one,
have been suggested.  Semeniuk et al. (1995) and Olech et al. (1996),
argued that it is the spin period of a rotating WD. Retter, Leibowitz
\& Ofek (1997) and Skillman et al. (1997) proposed, on the other hand,
that the longer period is that of PSH oscillations. We think that the
increase of this period during 1995 and 1996, as well as other features
of the optical LC of the nova discussed in Retter et al. (1997), argue
against the magnetic interpretation for the longer period, and for the
PSH one, which we adopt in this letter.

Two other old novae, V603 Aql 1918 (Patterson \& Richman 1991;
Patterson et al. 1993; Thomas 1993 and Patterson et al. 1997) and CP
Pup 1942 (White \& Honeycutt 1992; White, Honeycutt \& Horne 1993 and
Thomas 1993), also show PSH in their LCs.  While the presence of PSH in
the LC of V603 Aql is rather well established, there is still some
controversy in the case of CP Pup.  White et al. (1993) and Balman,
Orio \& Ogelman (1995) proposed a WD spin interpretation for the second
periodicity in the LC of this nova, which was initially thought to be
11\% longer than the orbital period. However, an extensive photometric
study by Thomas (1993) revealed that the period excess is only about
2\%, and that the two periods obey the well-known relation of Stolz \&
Schoembs (1984) between orbital and superhumps periods in SU UMa
systems. In addition, the spin periods in all but one intermediate
polars are shorter than their orbital periods (Patterson 1994). Even in
the one exceptional case (RX J19402-1025), the period excess is very
minute - $(P_{spin} - P_{orbital})/ P_{orbital} \approx$ 0.3\% (Patterson
et al. 1995), much smaller than in a typical superhump system.  We
believe that the observed photometric features of CP Pup favour the
superhump interpretation for this system and we adopt it in this work.

There are a few more reports on short periods in other novae.  RW UMi
1956, with a possible orbital period of $\sim$2~hr (Szkody et al.
1989), is a natural permanent superhump candidate, but intensive
continuous photometry of the nova during 25 nights in 1995 and 1997,
carried out at the Wise Obs., suggests that the variation is
irregular.  Two periods of $\sim$3.3~hr, which differ from each other by
less than 2\%, were found in V1500 Cyg 1975. These periods, however,
don't obey the Stolz \& Schoembs relation. The variation in the
polarization, found in this system, is also a very strong evidence for
the magnetic nature of the WD. (Stockman, Schmidt \& Lamb 1988).  There
are also some evidence that the 85-m period found in GQ Mus 1983 is
caused by the rotation of the magnetic pole of the primary (Diaz \&
Steiner 1994), thus cannot be identified with a superhump variation.

\section{Calculations}

\subsection{Deriving the Equations}

For each system, with a given orbital period, $P_{orb}$, in hours, the
critical mass transfer rate, that separates between the thermally
stable and unstable discs is given by Osaki (1996):

\begin{equation}
\dot{M}_{crit}\approx2.7\times 10^{17} (\frac{P_{orb}}{4hour})^{1.7}
\frac{gr}{sec.}
\end{equation}

The bolometric brightness of the disc is (Patterson 1994):

\begin{equation}
L_{bol}=\frac{GM_{wd}\dot{M}}{2R_{wd}}
\end{equation}

where G is the gravitational constant; $M_{wd}$ and $R_{wd}$ are the
mass and radius of the WD and $\dot{M}$ is the mass transfer rate.
For a 1.05 $M_{\odot}$ primary we take $R_{wd} \approx$ 5,500 km, and
$R_{wd} \sim$ $M_{wd}^{-1/3}$ (Carol \& Ostlie 1996). We also use the
basic equation (Allen 1976):

\begin{equation}
M_{bol}=4.75-2.5log(L_{bol}/L_{\odot})
\end{equation}

where $M_{bol}$ is the bolometric magnitude of the disc.
From equations 1,2 \& 3 we obtain the following relation:

\begin{equation}
M_{bol}\approx5.02-4.25log(P_{orb})-3.33log(M_{1})
\end{equation}

where $M_{1}$ is the mass of the primary in solar units, and we used
the mass-radius relation mentioned above for eliminating the radius.
For a given WD mass, $M_{1}$, equation 4 gives the bolometric magnitude
of a disc-dominated system crossing the tilted border line in Fig.~1,
as a function of the orbital period.

The bolometric magnitude is hard to assess observationally, because
observations at all wavelengths are needed. The ratio between the
visual luminosity of the accretion disc to its bolometric brightness
may, however, be estimated from theoretical models of stable accretion
discs. LaDous (1989) calculated such a model for a 1$M_{\odot}$ primary
with $\dot{M}\approx 1\times 10^{-9} M_{\odot}/yr$. From this model we
find:

\begin{equation}
\frac{L_{V}}{L_{bol}}\sim0.14.
\end{equation}

We note here that the main point presented in this letter is not very
sensitive to the exact value of this ratio. It does not change by much even
if there is an error of a factor two in its value.
We can now use the distance modulus equation (Allen 1976):

\begin{equation}
m_{V}=M_{V}-5+5log(d)+A_{V}
\end{equation}

where $m_{V}$ is the apparent visual magnitude of the disc; d is the
distance to the object in parsec and $A_{V}$ is the interstellar extinction
in the V band.
The final relation between the critical $m_{V}$ and the parameters of
the binary system is therefore:

\begin{equation}
(m_{V})_{crit}=2.16-4.25log(P_{orb})-3.33log(M_{1})+5log(d)+A_{V}
\end{equation}

We can also use the above equations in a different order and write down
an expression for the mass transfer rate as a function of the apparent
visual magnitude of the nova:

\begin{equation}
\dot{M_{17}}=(10^{\frac{m_{V}-A_{V}-0.69}{-2.5}})\frac{d^{2}}{M_{1}^{4/3}}
\end{equation}

where $\dot{M_{17}}=\dot{M}/(10^{17}gr/sec.)$. Deriving the former two
equations, we assumed that the visual brightness of the nova system
emanates solely from the accretion disc. This obviously neglects the
light of the secondary star in the system. Other light sources, such as
the nebula and the WD, probably contribute to the brightness of the
system as well. These contributions, however, diminish in time, as the
system decays after the nova event. We may also relax somewhat our
assumption of the exclusiveness of the accretion disc as a light
source, without changing considerably our final conclusion, except for
shortening the time required for the system to cross the critical
$\dot{M}$ line.

\subsection{V1974 Cyg}

Table 1 presents values of the input parameters to the equations of
the previous section for the three PSH classical novae. The ranges of
possible values given in the table are intervals between two estimates
of the corresponding parameter that were made in the cited literature.
For V1974 Cyg there are also a few estimates that are outside the
intervals given in the table. We shall discuss them in Section 3.5.

For Nova Cygni, with $P_{orb}\approx 1.95$ hr, equation 1 yields:
$\dot{M}_{crit}\approx 8\times 10^{16}$ gr/sec, and the range of
visual magnitudes that we extract from equation 7 is:
$m_{V}=17.9-18.5$. These numbers are much smaller than the pre-nova
magnitude of the system, which is either $m_{V}=19.5$ (Annuk, Kolka \&
Leedjarv 1993) or $m_{B}=21\pm1$ (Skiff 1997). It is, on the other
hand, still larger than the magnitude of V1974 Cyg in 1997 May -
$m_{V}\approx15.85$.

Using equation 8, with this visual magnitude of the nova, we derive a
mass transfer rate of $\dot{M}\approx 7\pm3 \times 10^{17}$ gr/sec.
This value is illustrated in Fig 1. (at the upper left domain), and it
is, reasonably, larger than the critical $\dot{M}$.

\begin{table}
\caption{Parameters of the Three Novae Exhibiting Permanent Superhumps}
\begin{tabular}{@{}lllcl@{}}

Object  & $P_{orb}$ & $M_{wd}$           &  d             & $A_{V}$ \\
        &    (hr)   &($M_{\odot}$)       &(kpc)           &         \\
\\

V1974 Cyg& 1.95$^1$ & 0.89-1.07$^{2,3,4}$&1.66-1.88$^5$   & 0.96-1.02$^{5}$\\
V603 Aql & 3.32$^6$ & 0.66-0.90$^{7}$    &0.33-0.38$^{7}$ &0.22-0.5 $^{8,9}$\\
CP Pup   &1.47$^{10}$&0.12-0.86$^{10,11}$&  0.83$^{12}$   &0.78-0.86$^{9,13}$\\

\end{tabular}
\end{table}
$^1$ DeYoung \& Schmidt 1994.\\
$^2$ Paresce et al. 1995.\\
$^3$ Balman, Krautter \& Ogelman 1997.\\
$^4$ Retter et al. 1997.\\
$^5$ Chochol et al. 1997.\\
$^6$ Patterson et al. 1993.\\
$^7$ Friedjung, Selvelli \& Cassatella 1997.\\
$^8$ Krautter et al. 1981.\\
$^9$ Warner 1995.\\
$^{10}$ Duerbeck, Seitter \& Duemmler 1987.\\
$^{11}$ White et al. 1993.\\
$^{12}$ Bode, Seaquist \& Evans 1987.\\
$^{13}$ Diaz \& Bruch 1997.\\

\subsection{V603 Aql and CP Pup}

Similar calculations can be made for the two other novae exhibiting
PSH in their LCs. The critical visual magnitude
values for V603 Aql are in the range $m_{V}=12.9-14.0$ and for CP Pup it is
$m_{V}=17.1-20.0$. The present visual magnitude of the two
systems are - $m_{V}\approx 12$ and $m_{V}\approx 15$, respectively
(Warner 1995). Both are appreciably brighter than the
corresponding critical values.

Using the current magnitudes of these two novae for calculating the
present mass transfer rate, we find $\dot{M}\approx 8\pm4 \times
10^{17}$ gr/sec for V603 Aql and $\dot{M}\approx 2.8\pm2.5 \times
10^{18}$ gr/sec for CP Pup.

\section{Discussion}

\subsection{Consistency}

The value of the parameters of the three nova systems that we used in
the previous section were derived by various researchers, in general
independently of the superhumps phenomenon. Our calculations show that
according to Osaki's theory, the very presence of superhumps in the LCs
of these systems is indeed expected from these values.

Further support for the validity of the calculations presented in this
work, particularly for the bolometric correction that we used in
Section 2.1 (equation 5), comes from the agreement of the mass
transfer rate that we obtain for V603 Aql with estimates by other
authors using other methods. We derive in Section 2.3 for this
parameter the value $8\pm4 \times 10^{17}$ gr/sec. It is in good
agreement with  $4.8 \times 10^{17}$ gr/sec (Krautter et al. 1981),
$7.6 \times 10^{17}$ gr/sec  (Wade 1988), and $2.6 \times 10^{17}$
gr/sec (Duerbeck 1992).

\subsection{The principal light source}

Our calculations in Section 2 were made with the assumption that the
visual light of all three novae emanates from the accretion discs in
these systems. This assumption is supported by the optical spectrum of
the three stars. Shai Kaspi kindly took for us two spectrograms of
V1974 Cyg with the FOSC camera at the Wise Obs., one in 1995 July and
one in 1996 August. The two spectra are dominated by nebular emission
lines and by a continuum that seems to be free of stellar absorption
features.  Similarly, spectra of V603 Aql (Williams 1983) and of CP Pup
(O'Donoghue et al. 1989, White et al. 1993), which were taken 69 years
after outburst in the V603 Aql case, and 43, 45 and 46 years after
outburst in the CP Pup case, do not show any obvious stellar absorption
features. Thus, any contribution of the secondary to the light of these
systems is indeed negligible.

Leibowitz (1993) has drawn attention to the appearance of a kink in the
visual LC of many classical novae a few tens of days after maximum
light. He suggested that the abrupt change in the slope of the decaying
LC signifies a transition of the main light source in the system from
the envelope of the contracting nova to the accretion disc in the
system.  In the two old novae, V603 Aql and CP Pup, the photometry
capable of detecting superhumps was performed many years after the
outburst, long after the appearance of the kink in their LCs. In V1974
Cyg, the superhumps were also detected only after the kink in the LC of
this nova, as shown in Fig.~2. This figure is a comprehensive visual
light curve of the nova, from outburst in 1992 February to 1997 May.
The data were taken from various amateur groups. The arrow in the
figure indicates the time when superhumps were first observed. Since
superhumps are a pure disc phenomenon and consistently with Leibowitz's
hypothesis we may conclude that presently in V1974 Cyg, the accretion
disc is indeed the major source of the optical continuum.

\subsection{The future of the permanent superhumps in the two old
novae}

We showed in Section 2 that in the three novae discussed in this
letter, the present visual magnitude is brighter than the critical
value for transition from the PSH to the SU UMa state. The future of
the superhumps in their LCs is correlated with the future run of their
visual LCs.

The question of what is the long term behaviour of the LC of classical
novae, many years after outburst, is a wide open one. It is difficult
to answer it observationally because of the scarcity of these objects
on one hand, and the early age, of less than a century of modern
observations in novae, on the other. This time duration is probably
still shorter than the characteristic time of the returning of
classical novae to their real quiescence state.

In two pioneering works, Vogt (1990) and Duerbeck (1992) attempted to
systematically investigate the long term photometric behaviour of
classical novae. Duerbeck's sample of 21 old novae includes also the
two old novae discussed in this letter. There are large systematic
difficulties and uncertainties in the determination of the decline rate
of old novae, as underlined by Duerbeck himself. He, nevertheless,
suggested for V603 Aql an average decline rate of 10.7 mmag/year, and
for CP Pup 36.9 mmag/year (Duerbeck 1992 Table 1).

Taking these numbers at face value and using our results of Section
2.3, we obtain that V603 Aql will cross the critical line in Fig. 1 not
before some 75 years in the future from today. CP Pup will undergo this
phase transition not before some 55 years from today. In view of the
large uncertainties in the value of the critical visual magnitude of
these two novae, as well as in their average decline rate, these time
intervals are rough lower limits at best.

\subsection{The future of the permanent superhumps in Nova Cygni 1992}

V1974 Cyg is still in a phase of relatively steep decline from its
recent outburst. In its visual LC shown in Fig. 2, three sections of
nearly linear decline, at three different slopes, are clearly
discernible. The decline rate in the last section, lasting now for more
than two  years, is about 0.7 mag/year. If the nova keeps this decline
rate for another few years, it will reach the critical visual
magnitude, indicated by the dot in the figure, sometime around the year
2000.  The vertical error bars around the dot denote the formal
uncertainty in the critical visual magnitude value (see Section 2.2).
The horizontal bars represent the uncertainty in the time of crossing
the line of phase transition, due to the uncertainty in the slope of
the present average linear decline.

\begin{figure}

\centerline{\epsfxsize=3.0in\epsfbox{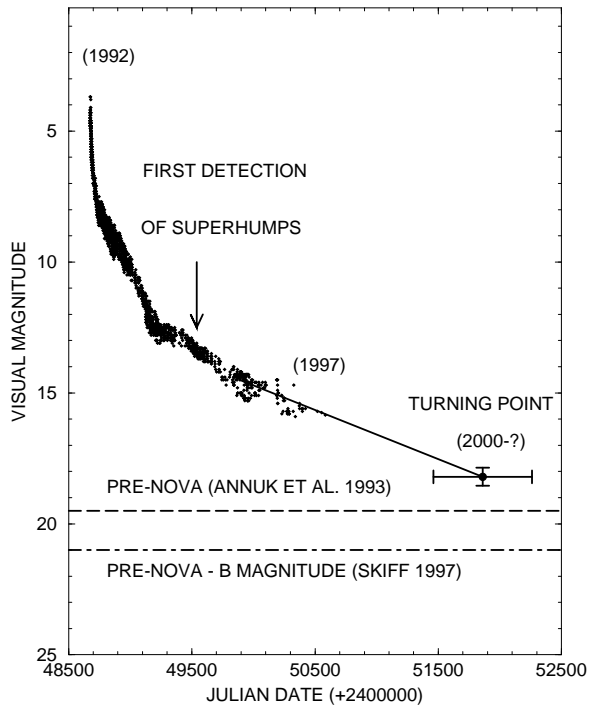}}

\caption{Five years light curve of Nova~Cyg~92. The data were taken
from visual estimates of three groups of amateur astronomers - AFOEV
(main source), VSNET and AAVSO. The solid line is a linear
extrapolation based on the mean decline rate of the nova in the last
two years. See text for further details.}

\end{figure}

\subsection{Sensitivity of the results to the adopted parameters
of N. Cygni 1992}

Table 1 presents values of system parameters that we used in our
calculations.  When more than one value is suggested in the literature,
the table gives a range of possible values, where the range limits are
two particular values that have been suggested for the corresponding
parameter. The results of our calculations in Section 2.2 and 2.3 are
accordingly given also as interval of possible critical values. For a
few parameters of V1974 Cyg, however, there are in the literature more
extreme estimates that put the parameter value outside the indicated
interval. Here we show that even when using these more extreme values
in our calculations, the results do not change very significantly.

Balman et al. (1997) suggested an upper limit of 1.37 $M_{\odot}$ on
the mass of the WD in V1974 Cyg. Paresce et al. (1995) proposed the low
value of 0.75 $M_{\odot}$ for this parameter. The use of these two
extreme values in our calculation widens the range of the predicted
critical visual magnitude of V1974 Cyg by no more than 0.4 magnitude on
each side. Similarly, the upper limit on the distance to this system of
3.2 kpc (Paresce et al. 1995) increases the critical range of
magnitudes by about 1.1 mag. It seems, however, that the expansion
velocities of the nebulae used by these authors are rather large. All
these possible modifications, while changing the calculated time that
is required before phase transition occurs, do not alter the
qualitative scenario, presented in this work. Nova Cyg '92 is indeed
expected to decline further in its optical brightness, as it is
presently still nearly four magnitudes brighter than the system
pre-nova magnitude $m_{V}=19.5$ (upper horizontal line in Fig. 2 -
Annuk et al. 1993). It may have even larger room to decline, if the
progenitor had the magnitude $m_{B}=21$ (Fig. 2 - lower horizontal line
- Skiff 1997).  Naturally, the possibility that V1974 Cyg changes its
rate of decline, or even reverses it into brightening, before reaching
the critical visual magnitude, cannot be ruled out. In this case PSH
will continue to prevail in the LC of this system, much like they do in
the LCs of V603 Aql and CP Pup.

\subsection{A proposed evolution scenario}

Based on the example of the three classical novae V603 Aql, CP Pup and
V1974 Cyg, and on the calculations presented in this letter, we may
speculate that short period novae with non-magnetic WDs may be the
progenitors of PSH systems. This quasi-stable stage in nova life might
take a few decades or centuries, before the system returns to its
quiescent state, and then becomes a regular SU UMa star. Naturally,
such a scenario must be checked quantitatively by a proper statistical
analysis of the population of the involved classes of stars.

\section{Acknowledgments}

We thank Y. Osaki, D. Prialnik, C. laDous, E. O. Ofek, T. Kato, O.
Shemmer and A. Gal-Yam for fruitful discussions and H. Duerbeck for
useful comments. In this research we have used and acknowledge with
thanks data from the AFOEV (Association Francaise des Observateurs
d'Etoiles Variables) database, operated at CDS, France, from the VSNET
(Variable Stars Net, Kyoto University, Japan) and from the AAVSO
(American Association of Variable Stars Observers) international
database, based on observations submitted to them by variable star
observers worldwide.  Astronomy at the Wise Observatory is supported by
grants from the Israeli Academy of Sciences.

\end{document}